# Exceeding the Nonlinear Shannon-Limit in Coherent Optical Communications by MIMO Machine Learning

Elias Giacoumidis[1], Jinlong Wei[2], Ivan Aldaya[3], and Liam P. Barry[1]

[1] Dublin City University, School of Electronic Engineering, Radio and Optical Communications Lab, Dublin 9, Ireland. elias.giacoumidis@dcu.ie

[2] Huawei Düsseldorf GmbH, European Research Center, Riesstrasse 25, München, Germany.

[3] Paulo State University, Campus Sao Joao da Boa Vista, 505 São João da Boa Vista, SP, Brazil.

**The nonlinear Shannon capacity limit has been identified as the fundamental barrier to the maximum rate of transmitted information in optical communications. In long-haul high-bandwidth optical networks, this limit is mainly attributed to deterministic Kerr-induced fiber nonlinearities and from the interaction of amplified spontaneous emission noise from cascaded optical amplifiers with fiber nonlinearity: the stochastic parametric noise amplification. Unlike earlier impractical approaches that compensate solely deterministic nonlinearities, here we demonstrate a novel electronic-based deep neural network with multiple-inputs and outputs (MIMO) that tackles the interplay of deterministic and stochastic nonlinearity manifestation in coherent optical signals. Our demonstration shows that MIMO deep learning can compensate nonlinear inter-carrier crosstalk effects even in the presence of frequency stochastic variations, which has hitherto been considered impossible. Our solution significantly outperforms conventional machine learning and gold-standard nonlinear equalizers without sacrificing computational complexity, leading to record-breaking transmission performance for up to 40 Gbit/sec high-spectral-efficient optical signals.**

Surging data traffic demands caused by Internet services such as video streaming or cloud computing pose a significant challenge to the underlying fiber-optic communication systems. To increase data rates, one of the core difficulties is the optical Kerr effect that arises at high signal powers causing a variation in index of refraction which is proportional to the local irradiance of the light[1] and is responsible for nonlinear optical effects such as self-phase modulation (SPM), cross-phase modulation (XPM), and four-wave mixing (FWM) that generates a fourth idler photon[2]. The optical Kerr effect is attributed to the so-called nonlinear Shannon capacity limit[2] which sets an upper bound on the achievable data rate in optical fiber communications when using traditional linear transmission techniques. There have been extensive efforts in attempting to surpass the nonlinear Shannon limit through several fiber nonlinearity compensation techniques and nonlinear transmission schemes[3–9] that compensate the deterministic Kerr-induced nonlinear effects. Albeit the Kerr-mediate nonlinear process is deterministic, the frequency uncertainty of many independent wavelength channels is transformed into time uncertainty through fiber transmission by chromatic dispersion, making the nonlinear interaction appear random[4]. On the other hand, high-capacity optical networks involve optical amplifiers (Erbium-doped fiber amplifiers, EDFAs) to compensate fiber losses, causing stochastic parametric noise amplification (PNA)[3] by means of the interplay between amplified spontaneous emission (ASE) noise (acting as Gaussian noise) and fiber nonlinearity. PNA is particularly debilitating for long-haul networks that utilize cascaded EDFAs, where the lower optical signal-to-noise ratio (OSNR) would prompt interest in nonlinearity compensation, since it scales at least quadratically with the system length[3]. Whilst the need for competitive edge ensures commercial interest in practically implementable form of nonlinearity compensation, it is difficult to imagine that a 50% increase in capacity would postpone an upcoming "capacity crunch" for the long-term[3].

The most advanced techniques that attempt to surpass the nonlinear Shannon limit, such as optical phase conjugation (OPC)[5], digital back-propagation (DBP)[4,6,7], phase-conjugated twin-waves (PC-TW)[8] and the most recent nonlinear Fourier transform (NFT)[8], compensate exclusively deterministic nonlinearities. However, these schemes are unable to tackle stochastic-induced nonlinearities such as PNA or the interaction of polarization-mode dispersion (PMD) with fiber nonlinearity. Additionally, OPC which is placed in the middle of a link significantly reduces the flexibility in an optically routed network requiring both symmetric chromatic dispersion map and power evolution, DBP and NFT are extremely complex for real-time signal processing, and PC-TW halves the transmission capacity by sacrificing one of the two polarizations with unused 'conjugated data'. Recently, machine learning has been under the spotlight for fiber nonlinearity compensation in digital domain[10]; harnessing various low-complex algorithms such as supervised artificial neural network (ANN)[11–13] and support vector machine (SVM)[14–18] based classifiers, with the ability to tackle both stochastic and deterministic sources of noise. Machine learning digital blocks are independent from mathematically tractable models and can be optimized for a specific hardware configuration and channel. Nevertheless, reported nonlinearity compensation via machine learning for long-haul coherent optical systems shows modest signal quality-factor (Q-factor) enhancement[11–18] since it cannot effectively compensate the PNA effect nor the deterministic FWM.

Recent advances in the field of machine learning have shown that deep neural networks[19,20] based on ANN classifiers have been rediscovered as a breakthrough technique for statistical learning in speech and image processing by attempting to simulate more effectively the human brain functionality. In deep learning, layer-wise training in multi-layer deep neural networks[21,22] is taken place with sufficient number of training data (also called epochs for the case of ANN). In this work, we demonstrate a novel MIMO deep learning based digital nonlinearity compensator for multi-gigabit per second (Gbit s$^{-1}$) coherent optical multi-carrier signals. In our proposed machine learning scheme, the hidden layers are adaptively adjusted by different sets of electronic frequency-subcarriers (Megahertz [MHz]-bandwidth per subcarrier) representing the third dimension, thus assisting information about the frequency location of the FWM nonlinear terms. We demonstrate that due to this attractive feature, adaptive MIMO deep learning outperforms conventional ANN[11] without considerably increasing computational complexity. MIMO deep learning results in record-breaking Q-factor enhancement of 3.8 dB compared to commercialized quaternary phase-shift keying (QPSK) single-carrier/polarization multi-channel optical signals of total 400 Gbit s$^{-1}$ signal capacity being transmitted at 3200 km of fiber length. Our approach reveals compatibility and effectiveness over future-proof high-order modulation formats such as 16-quadrature amplitude modulation (16-QAM). In comparison to DBP – which is perceived as the gold-standard digital nonlinearity compensation technique – and the conventional ANN, our approach can reduce deterministic nonlinear distortions and PNA by >2 and 4 dB, respectively.

**Principle of adaptive MIMO deep neural network**

Implementation of machine learning in telecommunications have a long history covering a wide range of applications, such as channel modeling and prediction, equalization, demodulation/modulation recognition, and spectrum sensing[23,24]. Deep learning is a branch of machine learning in which a neural network with multiple layers becomes sensitive to progressively more abstract patterns. Deep learning systems learn on their own from training data sets, until they can see patterns and spot very messy anomalies in data sets more effective than conventional machine learning and artificial intelligence. It is based on a set of algorithms that attempt to model high-level abstractions in data by using model architecture, with complex structures or otherwise, composed of multiple nonlinear transformations. The term "deep" refers to the number of sequential layers within a network. Depth relates directly to the number of iterative operations performed on input data through sequential layers' transfer functions. "Width" is used to describe the number of output activations per layer, or for all layers on average, and relates directly to the memory required by each layer. In this work, a novel MIMO deep learning based nonlinearity compensator harnessing ANN classification is implemented for coherent optical orthogonal frequency-division multiplexed (CO-OFDM)[25] signals, which consist of many frequency (electronic) subcarriers of MHz bandwidth. The reason for adopting the high-spectral-efficient OFDM over other multi-carrier schemes such as Nyquist-wavelength-division multiplexing (WDM)[26] is due to the insertion of a cyclic prefix (CP)[25] between symbols in time domain, which corrects the inter-symbol interference effect from the stochastic PMD in long-haul links without requiring additional digital signal processing (DSP) algorithms with heavy computational load[25]. In this manner, by virtually compensating PMD we truly investigate the impact of ASE noise at low transmitted powers; over which the received data reveal high entropy, meaning they have higher randomness due to EDFAs' non-deterministic noise.

The proposed high-performance equalizer based on adaptive MIMO deep learning is illustrated in the conceptual diagram of Fig. 1. In a coherent optical communication system, the in-phase (I) and quadrature (Q) components are detected, digitized, and combined in electronic domain to build a single complex signal that retains the information of both the optical amplitude and phase. In a multi-span high-capacity multi-carrier optical system such as CO-OFDM, this signal is corrupted by the following three effects: (i) Deterministic nonlinear distortions arising from the inter-carrier interference (ICI) induced by the Kerr effect through FWM. Given the square shape of OFDM spectrum and the degeneracy of FWM process, deterministic nonlinear distortion has a characteristic triangular shape as shown in Fig. 1. (ii) Additive noise is stochastic and, in coherent multi-span systems, is dominated by the ASE of the optical amplifiers. In the signal bandwidth, this noise has a flat power density. (iii) The nonlinear interaction between the information OFDM signal and ASE noise in the several spans gives rise to the so-called parametrically amplified noise (or PNA[3]), a term that indicates that the ASE noise is amplified via parametric process where the OFDM signal plays the role of the pump. This effect is also stochastic and flat over the signal bandwidth. The combined effect of the different degradation mechanisms can be assessed through the signal-to-interference-plus-noise ratio (SINR). It should be noted that inter-channel nonlinearities in WDM-CO-OFDM affect the edge subcarriers' SINR only, however, due to a sufficient channel spacing out-of-band nonlinearities are not as severe as intra-channel nonlinearities.

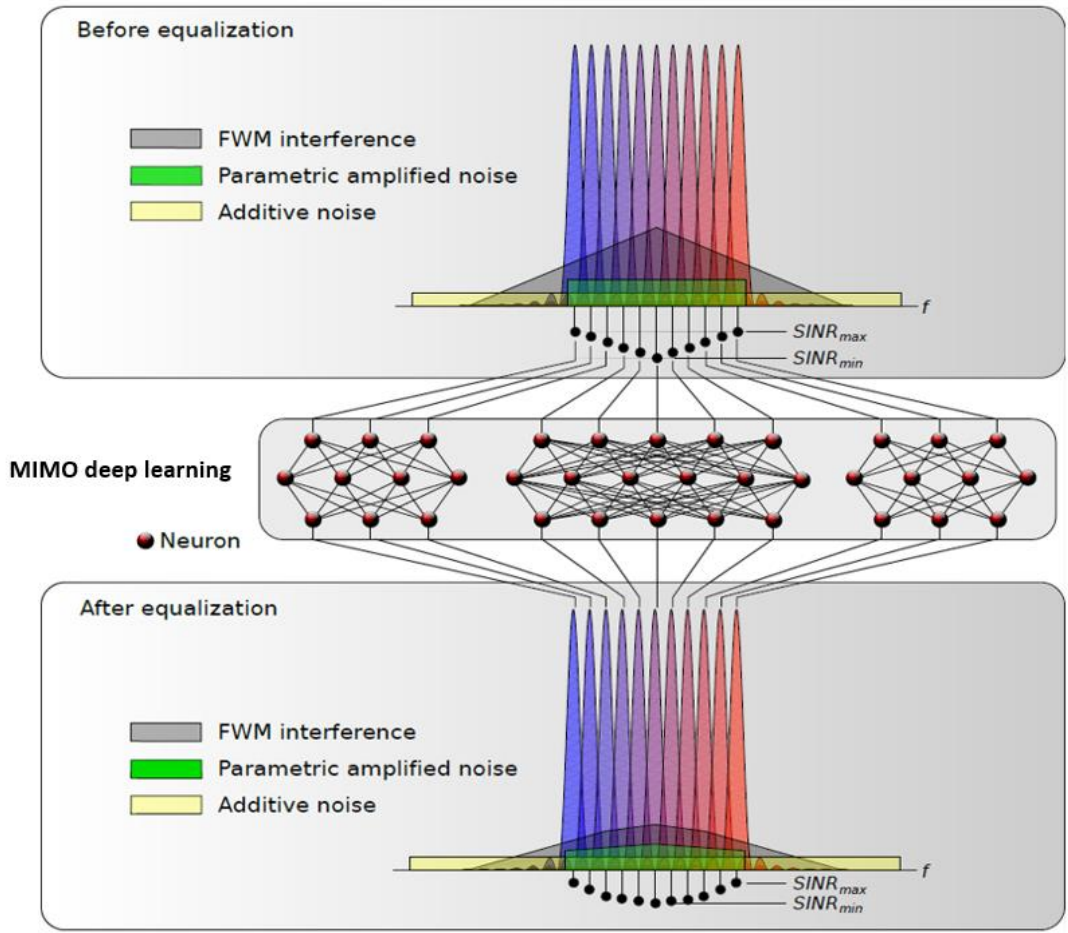

**Figure 1 | Conceptual diagram of the impact of the proposed adaptive artificial neural network (ANN) based MIMO deep learning equalizer on multi-carrier based coherent optical OFDM (CO-OFDM) signals.** The impact of the proposed nonlinear equalizer on the four-wave mixing (FWM) interference and the parametric amplified noise in illustrated. SINR, signal-to-interference-plus-noise ratio; f, frequency.

As sketched in Fig. 1, even if additive and parametric amplified noises are flat, the frequency-dependent deterministic distortion leads to a different SINR for each subcarrier, with the central subcarriers presenting a lower SINR ($SINR_{min}$) than the edge subcarriers ($SINR_{max}$). Consequently, we developed a nonlinear equalizer based on ANN that employs a higher number of interconnected neurons (also called perceptrons) to process subcarriers that are critically affected. As schematically shown in Fig. 1, the central subcarriers are processed with a more complex ANN compared to those located on the edges of the OFDM signal. Following this strategy, we can reduce the effect of nonlinear distortion mainly in the most affected central subcarriers. The output SINR, however, may not be flat but the range between maximum and minimum is reduced, leading to a more robust equalization and an overall performance improvement.

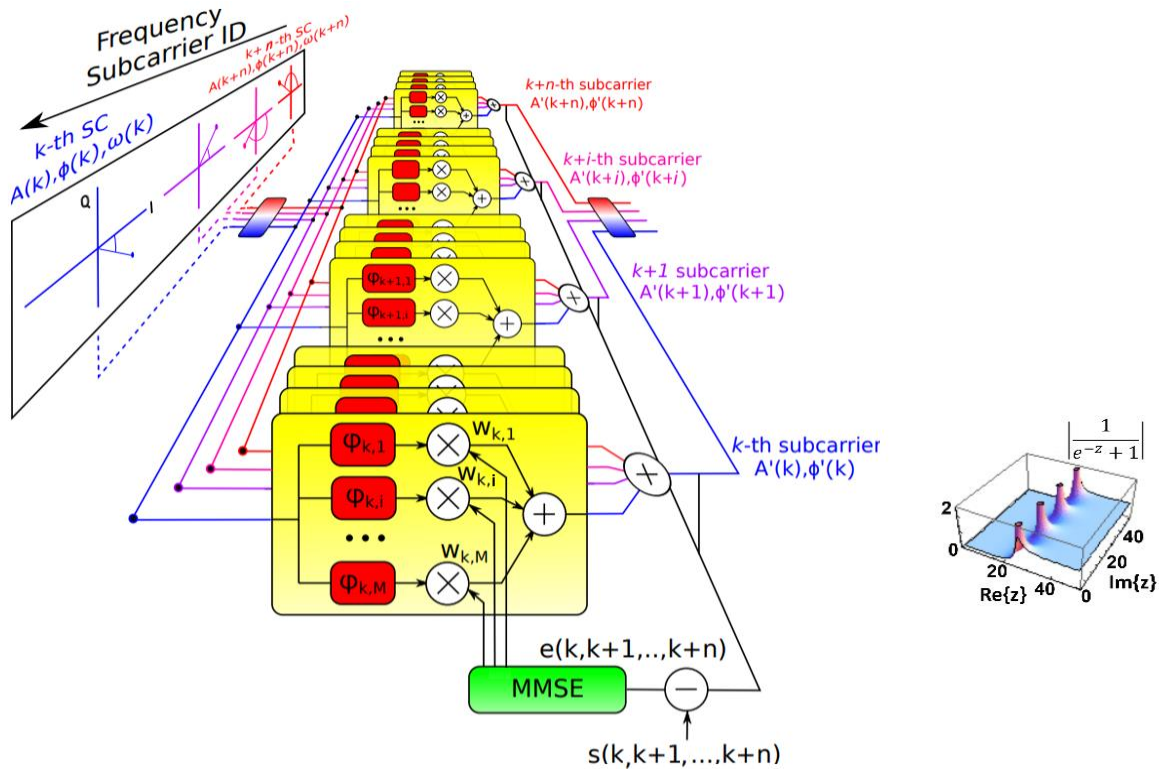

**Figure 2 | Block diagram of MIMO deep learning's centre hidden layer for middle CO-OFDM subcarriers.** The proposed nonlinear equalizer processes amplitude (A), phase (Φ) and frequency (angular, ω) at once. The block diagram shows the combination of random complex symbols from four different subcarriers (k, k+1, k+$i$-th, and k+n) at a distinct frequency. Inset: Example of a sigmoid function response. ID, identifier/index; SC, subcarrier; I, in-phase; Q, quadrature; k, subcarrier index; M, modulation format level; φ, nonlinear mapping for phase-shift keying (PSK)/quadrature amplitude modulation (QAM) constellation diagrams; e, error; w, weight values (Lagrange multipliers); MMSE, minimum mean square error; s, desire output vector; A', output amplitude; Φ', output phase.

In Fig. 2, a block diagram of the proposed nonlinear equalizer is depicted for the centre hidden-layer in which amplitude, phase and frequency information of the middle CO-OFDM subcarriers is managed at once. The block diagram shows random multiplexed input complex symbols from four different subcarriers (k, k+1, k+i, and k+n) at a distinct frequency. The s term presents the training vector, i.e. the pre-known subcarrier/symbols set transmitted during the training stage. This could be high consuming in terms of required computational resources, nonetheless, for a highly stable system such as long-haul optical links in which chromatic dispersion and nonlinear effects do not change over time, the mapping could be accomplished once. The centre hidden layer of the MIMO deep learning nonlinear equalizer is comprised of up to k+n sub-neural networks coming from k+n subcarriers. The received symbols in every hidden-layer are fed by neurons which are multiplied with the weight value $w_{k,i}$ per subcarrier. Afterwards, the outputs from different neurons are summed to generate

ŝ(k,k+1,k+i,k+n); an estimation which minimizes the error between the transmitted and received symbols with the assistance of the training stage.

In the training stage, the minimum mean square error (MMSE) algorithm is adopted via a complex version of the Riedmiller's resilient back-propagation (RR-BP)[27,28] algorithm to determine the error signal and update the weights. The weights are iteratively updated until the desired error value is reached, thus indicating the optimum match between the sub-network output and the transmitted (undistorted) OFDM subcarrier symbols. The error signal per subcarrier/layer in MIMO deep learning is given by,

$$e(k) = s(k) - ŝ(k) \tag{1}$$

ŝ is calculated in terms of a nonlinear activation function φ which is considered the "heart" of the proposed nonlinear equalizer[11] and is given per subcarrier/layer by:

$$ŝ(k) = \sum_{i=1}^{M} w_{k,i} \varphi_{k,i} s(k). \tag{2}$$

The nonlinear activation function is application-dependent and it is mostly required to be a differentiable function. In our system design a sigmoid function was employed, which can satisfy a conflicting relationship between the boundedness and the differentiability of a complex function as shown in the right side inset of Fig. 2. Another difference with the conventional ANN design of[11–13] in which only the quadrature components are managed by an activation function, here both in-phase and quadrature components are processed thus accounting for their cross-information. The complexity of the proposed adaptive neural network is dependent on the number of hidden layers and the modulation format level, M. In this work, the number of hidden layers is either three of four while M is four or sixteen for QPSK and 16-QAM, respectively. More information about the MIMO deep learning is discussed in Methods.

In this work we considered four different cases for adaptive nonlinear equalization using MIMO deep learning by selecting distinct block-sizes for middle subcarriers and corresponding neurons interconnections which suffer the most from FWM[12]. For a total number of 210 generated subcarriers we selected three cases (Cases 1–3) with three groups of subcarriers and hidden layers. For the fourth case we selected four groups of subcarriers and corresponding hidden layers. The total number of neurons for all cases was estimated at 2,520 and 10,080 for QPSK and 16-QAM, respectively. In summary, the following cases in Table I were adopted:

**Table I. Adopted cases of block-sizes for middle subcarriers and number of hidden layers**

| Cases | Block-size for middle subcarriers | Hidden layers |
|---|---|---|
| Case 1 | 50 | 3 |
| Case 2 | 100 | 3 |
| Case 3 | 150 | 3 |
| Case 4 | 2×54 (middle groups); 2×51 (edge groups) | 4 |

**Benchmark machine learning and deterministic nonlinearity compensators**

Various benchmark machine learning algorithms have been recently introduced in digital domain for fiber nonlinearity compensation in coherent optical transmission systems. For instance, ANN and SVM have been implemented for supervised (training data dependent) nonlinear equalization, whereas the unsupervised K-means[29,30], Gaussian mixture[31], Hierarchical[32,33] and Fuzzy-logic C-means[33] clustering have been applied for blind nonlinearity compensation. For fair comparison, in this work we implement an ANN for CO-OFDM with procedure identical to[11,12] using 210 subcarriers on a one-by-one subcarrier basis. The DBP nonlinearity compensator was also adopted, which is widely known as the deterministic gold-standard algorithm in digital domain that solves the nonlinear Schrödinger equation using the split-step method to undo the combined effects of chromatic dispersion and fiber nonlinearity[14,15]. In this work, a full-step DBP was adopted taking into account 40 steps-per-span, above which similar performance is observed[34].

**Experimental transmission setups**

For our experiments (detailed in Methods), QPSK WDM-CO-OFDM and 16-QAM single-channel CO-OFDM were transmitted in one polarization of a single-mode fibre at 3200 and 2000 km, respectively, as depicted in Fig. 3. Whereas a single-channel system as shown in Fig. 3(a) generates only intra-channel nonlinearities such as SPM, a WDM configuration (Fig. 3(b)) also produces at the receiver-end inter-channel nonlinearities, i.e. FWM and XPM[14]. It should be noted that since a multi-carrier CO-OFDM signal is considered for single-channel transmission, this also produces intra-channel ICI FWM and XPM, with the first being more severe[11,12].

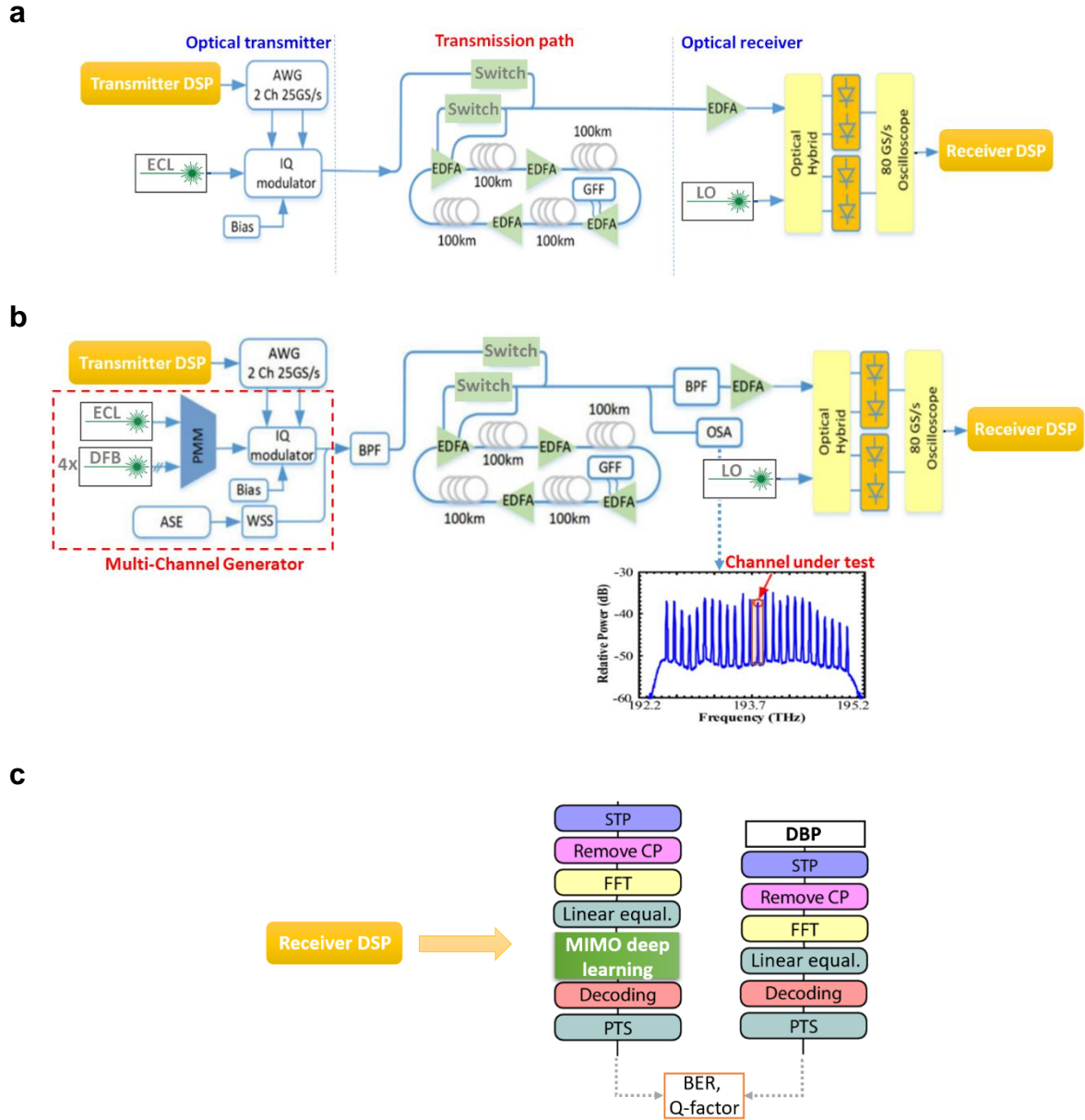

**Figure 3 | Experimental transmission set-up and DSP receiver design:** (**a**) Single-channel 16-QAM CO-OFDM system for 2000 km test. (**b**) WDM QPSK CO-OFDM for 3200 km test (middle-channel under test). (**c**) Receiver CO-OFDM-based DSP equipped with adaptive MIMO deep learning, ANN or DBP. AWG, arbitrary waveform generator; DSP, digital signal processing; ECL, external cavity laser; EDFA, Erbium-doped fiber amplifier; GFF, gain-flattening filter; LO, local oscillator; OSA, optical spectrum analyser; BPF, band-pass filter; DFB, distributed feedback filter; ASE, amplified spontaneous emission; WSS, wavelength-selective switch; PMM, polarization-maintaining multiplexer; STP/PTS, serial-to-parallel/parallel-to-serial; CP, cyclic prefix; FFT, fast Fourier transform; BER, bit-error-rate; Q-factor, quality-factor.

In the WDM configuration, 20 channels separated in frequency domain by 10 GHz (estimated from the centre of each channel) were loaded, covering 2.5 Terahertz of bandwidth as depicted in inset of Fig. 3(b) and thus carrying ~400 Gbit s$^{-1}$. The middle-channel suffering the most from inter-channel nonlinear crosstalk effects[12] was hereafter under test at the receiver-side of the WDM-CO-OFDM system. A 2% of CP length was inserted in OFDM to virtually eliminate PMD. For MIMO deep learning, ANN, DBP, and linear equalization the net signal bit-rate for the QPSK WDM-CO-OFDM system was fixed for each channel at 18.2 Gbit s$^{-1}$ after CP is removed (shown in Fig. 3(c)), and 16.84 Gbit s$^{-1}$ after 10% of training overhead for ANN and MIMO deep learning is removed (which as shown in results-part is the optimum training overhead), while the raw bit-rate was ~20 Gbit s$^{-1}$ per channel (total transmitted signal capacity of ~400 Gbit s$^{-1}$). For the 16-QAM single-channel CO-OFDM system the net and raw signal bit-rates were set at ~40 Gbit s$^{-1}$ (after machine learning overhead) and ~46 Gbit s$^{-1}$, respectively. The key experimental parameters are summarized in Table II in Methods. The OFDM receiver included carrier frequency offset estimation, chromatic dispersion compensation and pilot-assisted channel estimation[25]. The proposed and benchmark machine learning based nonlinear equalizers were placed after the fast Fourier transform (FFT) block of a typical OFDM digital receiver in frequency domain as shown in Fig. 3(c). On the other hand, the benchmark DBP was processed just before serial-to-parallel conversion. The total and individual OFDM subcarriers' bit-error-rate (BER) by error counting and Q-factor (=20log$_{10}[\sqrt{2}erfc^{-1}(2BER)]$) are the crucial parameters under investigation. More details on experimental setups are provided in Methods.

**Results**

In Fig. 4, the transmission performance of the proposed adaptive MIMO deep learning nonlinear equalizer is depicted in terms of launched optical power (LOP) using the four different cases for subcarrier group selection. All cases for nonlinear equalization were tested over a QPSK WDM-CO-OFDM system at 3200 km and a single-channel 16-QAM CO-OFDM at 2000 km, with results depicted in Figs. 4(a), (b). Firstly, it is shown that MIMO deep learning can enhance the Q-factor by 4 dB when compared to linear equalization in QPSK WDM-CO-WDM at optimum –5 dBm of LOP, and 5 dB for the 16-QAM single-channel system at optimum LOPs. Secondly, it is shown that for both single- and multi-channel configurations, Case 2 − nonlinear equalization with 100 middle subcarriers − outperforms the other three cases (50, 150, or 2×54 middle subcarriers) and linear equalization. The performance enhancement of Case 2 is higher for the WDM transmission system, since it can tackle more effectively all combined inter-channel nonlinearities, intra-channel ICI FWM and PNA. The ICI improvement using Case 2 shows that 100 centre subcarriers is identified as the optimum selection for effectively combating the impact of FWM (and thus increasing the total subcarriers' SINR), above which leads to overfitting and performance degradation. This is corroborated in Figs. 4(c), (d), where the subcarrier and symbol

Q-factor distribution is plotted at optimum LOP for Cases 1 and 2, respectively. As shown from these figures, the Q-factors on centre subcarriers for Case 2 clearly outperform to these of Case 1. Moreover, Fig. 4(a) shows that Case 2 of MIMO deep learning leads in a record-breaking Q-factor enhancement of 3.8 dB compared to a commercialized single-carrier/polarization QPSK system which[35] was simulated for a multi-channel configuration at 3200 km (details on simulated setup is provided in Methods). Such performance enhancement is maintained for 16-QAM modulation as shown in the simulated blue curve of Fig. 4(b).

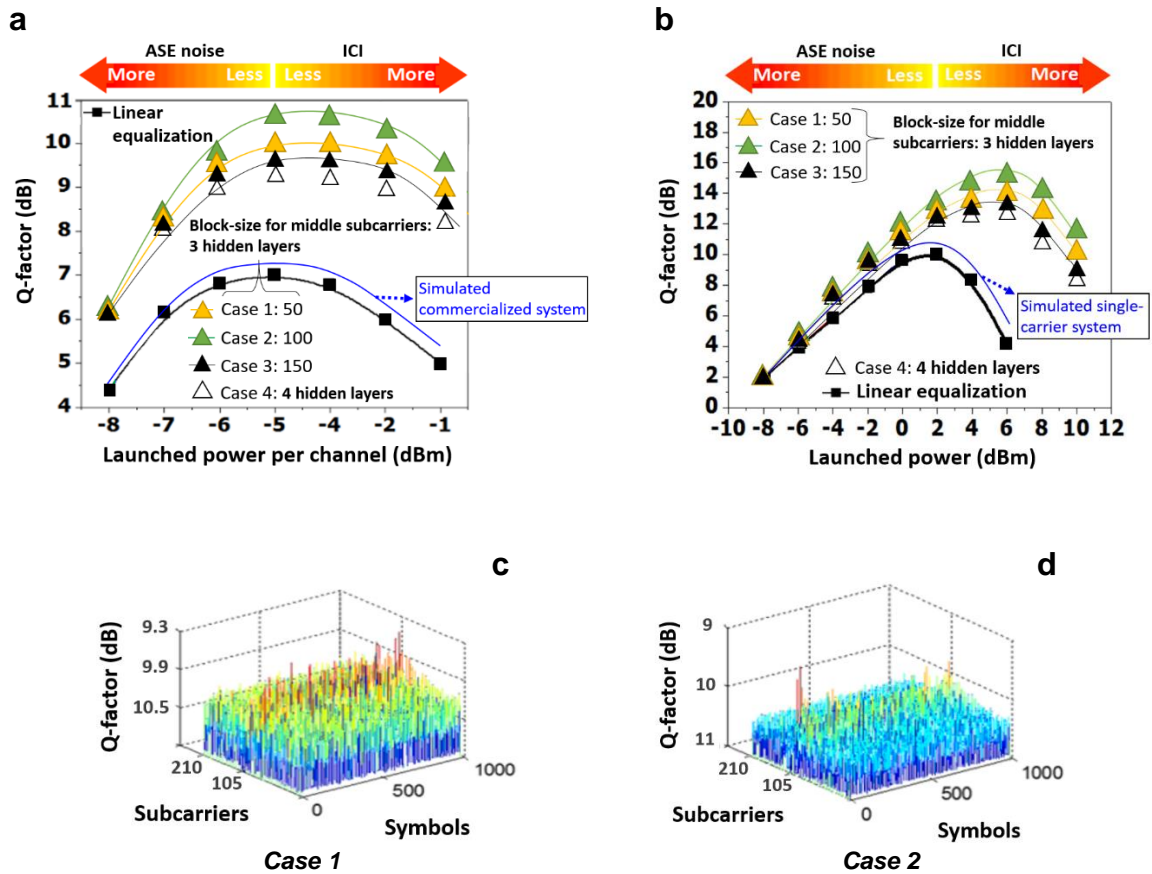

**Figure 4 | Transmission performance of adaptive MIMO deep learning using different block sizes for the equalization of the middle subcarriers.** Comparison of three cases with 50, 100 and 150 middle subcarriers that employ 3 hidden layers, and a fourth case in which 4 hidden layers are adopted with grouping of 2×54 (middle groups) and 2×51 subcarriers (edge groups) for: (**a**) WDM QPSK CO-OFDM at 3200 km (middle-channel under test); and (**b**) single-channel 16-QAM CO-OFDM at 2000 km. Blue curves indicate a simulated single-carrier QPSK (commercialized) and 16-QAM system. The Q-factor distribution on subcarriers/symbols in WDM configuration is shown at optimum –5 dBm of launched optical power (LOP) for: (**c**) Case 1: 50 centre subcarriers; and (**d**) Case 2: 100 centre subcarriers. ICI, inter-carrier interference.

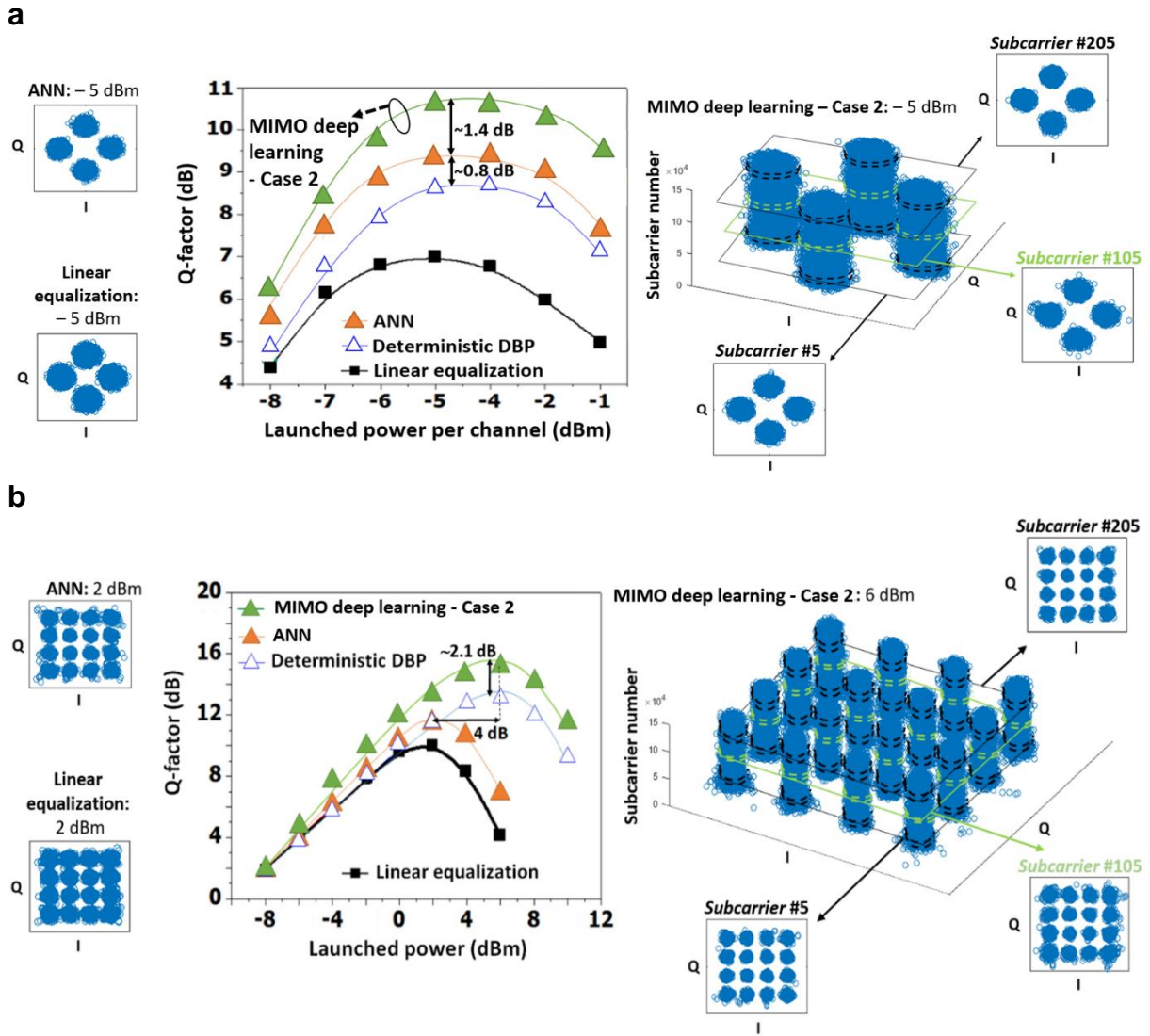

**Figure 5 | Transmission performance comparison between adaptive MIMO deep learning – Case 2 (3 hidden layers with 100 middle-layer subcarriers), ANN (210 subcarriers) and deterministic DBP.** Comparison of nonlinearity compensators in (**a**) WDM QPSK CO-OFDM at 3200 km (middle-channel under test); and (**b**) single-channel 16-QAM CO-OFDM at 2000 km.

In Fig. 5, we compare the best case of MIMO deep learning (i.e. Case 2) with the benchmark ANN and the deterministic DBP for (a) WDM QPSK CO-OFDM at 3200 km and (b) single-channel 16-QAM CO-OFDM at 2000 km. It is shown that at the optimum LOP of –5 dBm related to the WDM transmission system in Fig. 5(a), a Q-factor improvement is observed compared to ANN and DBP of ~1.4 and ~2.1, respectively. The remarkable improvement of our nonlinear equalizer in the WDM configuration over DBP is due to the fundamental inability of DBP to tackle WDM and multi-carrier nonlinear crosstalk effects without an accurate knowledge of the absolute position and the relative separation of both the channels and subcarrier frequencies[4]. In Fig. 5(b), results reveal an extension of 4 dB in the maximum allowed LOP when compared to linear equalization and ANN, and ~2.1 dB improvement in Q-factor in comparison to DBP at the optimum LOP of 6 dBm. In insets of Figs. 5(a), (b), the received QPSK/16QAM constellation diagrams are depicted at optimum LOPs, verifying that MIMO deep learning can correct the phase rotation of symbols (especially in 16-QAM) be means of nonlinear phase noise reduction.

Finally, the impact of the machine learning algorithms' training overhead on the transmission performance of the single- and multi-channel CO-OFDM is illustrated in Fig. 6. In this comparison we assume the best Case 2 for middle hidden layer subcarriers in MIMO deep learning. It is shown that above 10% of training overhead, the performance of the two nonlinear equalizers is saturated for both QPSK WDM-CO-OFDM and single-channel 16-QAM CO-OFDM. The reason that the Q-factor is degraded below 10% is due to underfitting, since less training samples lead to wrong classification because the small-size training data are more susceptible to noise. Hereafter, a 10% of training data was adopted throughout this work that clearly maximizes the Q-factor.

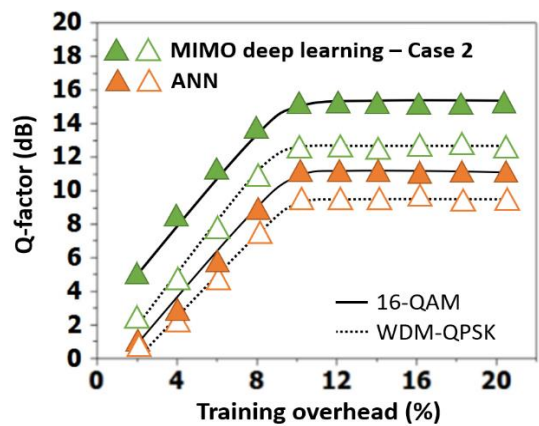

**Figure 6 | Training overhead evolution comparison between MIMO deep learning (Case 2) and ANN**. Comparison of single-channel 16-QAM CO-OFDM at 2000 km and WDM (middle-channel) QPSK CO-OFDM at 3200 km. The LOP for 16-QAM was selected at optimum 6 dBm for MIMO deep learning and 2 dBm for ANN. The LOP per channel for WDM-QPSK is –5 dBm for both nonlinearity compensators.

## Discussion

We have demonstrated the first MIMO deep learning based digital nonlinearity compensator for high-capacity coherent optical communications. Our demonstration using a novel adaptive implementation of MIMO deep learning resulted in compensation of both intra- and inter-channel nonlinear crosstalk effects in high-spectral-efficient optical signals even under the presence of frequency stochastic variations, which has never been addressed in the past. Compared to conventional ANN classifiers, our approach employed a higher number of neurons to process electronic-based frequency subcarriers that are critically distorted, avoiding underfitting and thus reducing the effect of nonlinear distortion mainly in the more affected central subcarriers. The proposed neural network design emulated more effectively the human brain activity and therefore was able to combat better residual FWM terms compared to ANN without adding computational complexity in both single- and multi-channel coherent optical systems. Our approach resulted in record-breaking signal Q-factor enhancement of 3.8 dB compared to commercialized single-carrier/polarization systems for ~400 Gbit s$^{-1}$ optical links at 3200 km of transmission. In comparison to the gold-standard deterministic DBP, MIMO deep learning revealed a transmission performance benefit by tackling the stochastic PNA effect even when using advanced modulation formats, concurrently providing a significantly lower computational effort[11]. MIMO deep learning essentially identified the optimum number of neurons, above which resulted to overfitting. It should be indicated that the nonlinear distortions in CO-OFDM are primary induced from FWM and PNA, and secondary from XPM. However, the SINR improvement on middle subcarriers is also a result of XPM reduction.

MIMO deep learning can seamlessly operate for low-cost direct-detected systems and dual-polarization signals using an extra 2×2 multiple-input multiple-output[36] digital block of butterfly structure to unravel the polarization crosstalk. The proposed nonlinear equalizer is also transparent to alternative multi-carrier optical schemes such as Nyquist-WDM[26] and filter-bank modulation[37]. It should be also noted that the high peak-to-average power ratio of the CO-OFDM signal made it more vulnerable to fiber nonlinearity. This produced significant in-band nonlinear distortion and spectrum spreading, causing the subcarrier intra-channel crosstalk effects appearing more random rather deterministic[15,19]. Hence, our demonstrations have shown the necessity of implementing deep learning based nonlinear equalization in CO-OFDM.

Throughout this work, digital-based MIMO deep learning has revealed the power of merging electronics and photonics with machine learning. The training process of our proposed scheme replaced the need for accurate channel information; and since it is of relatively simple mathematical structure makes it implementable for future high-speed DSPs[38]. This can potentially revolutionize real-time[39], low-latency telecommunications such as the Internet-of-Things, telemedicine, interactive entertainment and financial trading services, whereby saving a fraction of millisecond can be of pivotal importance[40,41]. Due to the potential of machine learning to tackle stochastic noises, the proposed solution could be possibly implemented in the DSP units of 5G mobile networks[42] (in which OFDM has already been standardized: IEEE 802.11a WiFi and 4G Long Term Evolution), visible-light[43,44] and satellite communication[45] systems for signal quality improvement. Finally, MIMO deep learning could trigger the introduction of fast neuromorphic digital mobile-phone chips[46] to handle sensory data and tasks such as image recognition[47].

**Methods**
**Experimental setups.** In the transmitter, the single- and multi-channel CO-OFDM systems employed external cavity lasers (ECLs) of 100 KHz linewidth, being modulated using a dual-parallel Mach-Zehnder modulator fed with 'offline' OFDM I-Q components. Single-polarization CO-OFDM was also considered for both single- and multi-channel configurations. The transmission path at 1550.2 nm was a recirculating loop consisting of 20×100 km (single-channel configuration) and 32×100 km (WDM configuration) spans of Sterlite OH-LITE fiber (attenuation of 18.9-19.5 dB/100 km) controlled by acousto-optic modulator. The loop switch was located in the mid-stage of the 1$^{st}$ EDFA and a gain-flattening filter was placed in the mid-stage of the 3$^{rd}$ EDFA for both configurations. For WDM CO-OFDM, four distributed feedback lasers and an ECL were employed on 100 GHz grid located between 193.5–193.9 THz. Using an ASE source, another 20 'dummy' WDM channels of 10 GHz bandwidth were generated. These channels covered 2.5 THz of bandwidth as depicted in the inset of Fig. 3(b). In the experiments, the baseband waveform samples were calculated offline based on a pseudo-random binary sequence of $2^{19-1}$. In the transmitter, an arbitrary waveform generator was used at a sampling rate (bandwidth) of 34 GHz (2 channels with 25 GSamples s$^{-1}$ for I and Q components) to generate a continuous baseband signal for both single-channel and WDM (middle-channel) CO-OFDM. At the receiver, for both single- and multi-channel systems, the optical receiver was constituted by a homodyne coherent detector with 100 kHz linewidth local oscillator laser (ECL-based). An oscilloscope was used at 80 GSamples s$^{-1}$ in real-time and the digital part was performed offline in Matlab®. 210 subcarriers were generated in a 512 inverse-FFT (IFFT) size using QPSK (WDM) and 16-QAM (single-channel) formats.

**Table II. Key experimental transceiver and transmission parameters.**

| Experimental parameters | Single-channel CO-OFDM | WDM-CO-OFDM |
|---|---|---|
| Modulation format | 16-QAM | QPSK |
| Raw signal bit-rate | ~46 Gbit s$^{-1}$ | ~20 Gbit s$^{-1}$ |
| Number of CO-OFDM channels | 1 | 20 |
| Total bandwidth (channel spacing) | 34 GHz (–) | 2.5 THz (10 GHz) |
| Transmission-reach | 2000 km [20×100 km spans] | 3200 km [32×100 km spans] |
| OFDM cyclic prefix length | 2 % | 2 % |
| ECL/DFB linewidth | 100 kHz | 100 kHz |
| Generated subcarriers (IFFT size) | 210 (512) | 210 (512) |

**MIMO deep learning algorithm.** The training function in MIMO deep learning updates the weights and the bias values using a complex RR-BP, performing an approximation to the global minimization achieved by the steepest descent[28]. For the general case of a vector n, RR-BP minimizes the difference between the MIMO deep learning output and the targeted output in each hidden layer using (3).

$$E(n) = \|s(n) - \hat{s}(n)\|^2 \quad (3)$$

where s(n) and $E(n) = \|s(n) - \hat{s}(n)\|^2$ are the desired and calculated output vectors, respectively. The employed transfer functions for the hidden layers of the MIMO deep learning were differentiable and similar to the hyperbolic tangent function, as suggested in[13]. For the output layer, the linear function "purelin" was used. The block identified as MMSE in Fig. 2, represents the subsystem that implements the RR-BP algorithm used to find the weights via Lagrange multipliers that minimizes the error vector (taken from (1)). The training function that updates the weights and bias values with RR-BP splits the complex OFDM data in two real-valued data collections. The weights are updated according to the following 5 steps by applying the gradient descent on the cost function E(n) in order to reach a minimum:

Step 1: Initialize the weights and thresholds to small random numbers.

Step 2: Present the input vector, X(n), and the desired output vector S(n).

Step 3: Calculate $\tilde{S}(n)$ from the X(n) and compute the error vector E(n) using (3).

Step 4: Adapt weights based on:

$$(w_{k,i})_{n+1} = (w_{k,i})_n - ñ(\partial E(n)/\partial w_{k,i} - n) \qquad (4)$$

where $(w_{k,i})_n$ is the weight of the i-th neuron of the k-th sub-neural network per hidden layer (i-th symbol of the k-th total number of subcarriers per hidden layer) at the n-th iteration, and ñ is the learning rate parameter. When ñ is very small, the algorithm will take long time to converge; whereas, when ñ is too large the system may run into an unstable state.

Step 5: If E(n) is above the threshold, go to step 2.

**Simulation setup**. The developed single-carrier simulated optical system was implemented in a Matlab/Virtual Photonics Inc. (VPI)-transmission-Maker® co-simulated environment (electrical domain in Matlab and optical components with standard single-mode fiber in VPI). The analogue-to-digital/digital-to-analogue converters clipping ratio and quantization have been considered and set at optimum 13 dB and 10-bits. For the in-line optical amplification, EDFAs were adopted having with 5.5 dB of noise figure. The EDFA noise was also modelled as additive white-Gaussian noise. The optical fiber for single-polarization transmission was modelled using the pseudo-spectral split-step Fourier method which solves the nonlinear Schrödinger equation. The adopted standard single-mode fiber parameters in this work are the following: fiber nonlinear Kerr parameter, chromatic dispersion, chromatic dispersion-slope, fiber loss, and polarization-mode dispersion coefficient of 1.1 $W^{-1}km^{-1}$, 16 ps $nm^{-1}km^{-1}$, 0.06 ps $km^{-1}$ $(nm^2)^{-1}$, 0.2 dB $km^{-1}$ and 0.1 ps $(km^{0.5})^{-1}$, respectively. The signal bit-rate was set to the exact same value to the experimental CO-OFDM for both QPSK and 16-QAM formats. Single single-polarization was used, the receiver digital receiver was based on blind adaptive equalization, and more particularly on the constant modulus algorithm which carries out residual chromatic dispersion compensation[48]. Frequency offset compensation and carrier phase estimation was done by the methods described in[49]. To avoid cycle slips, the detection was differential (i.e. DQPSK).